\documentclass{article}

\usepackage{amsmath,amssymb,theorem,epsfig}

\newcommand{\bp}{\boldsymbol{p}}

\newcommand{\bk}{\boldsymbol{k}}

\newcommand{\bbeta}{\boldsymbol{\beta}}
\newcommand{\uk}{\underline{k}}

\newcommand{\ua}{\underline{a}}
\newcommand{\ue}{\underline{e}}
\newcommand{\ub}{\underline{b}}
\newcommand{\ux}{\underline{x}}
\newcommand{\uba}{\underline{\boldsymbol{a}}}
\newcommand{\ubb}{\underline{\boldsymbol{b}}}
\newcommand{\ubk}{\underline{\boldsymbol{k}}}

\newcommand{\uxi}{\protect{\underline{\xi}}}

\newcommand{\bq}{\boldsymbol{q}}
\newcommand{\bx}{\boldsymbol{x}}
\newcommand{\qed}{\hfill$\square$}
\newcommand{\id}{\protect{\textrm{id}}}

\newcommand{\thh}[1]{$#1^{\text{th}}$}
\newcommand{\wick}[1]{\protect{{\boldsymbol{:}\!#1\!\boldsymbol{:}}}}

\newcommand{\Leff}{\mathcal L_{\text{eff}}}

\newtheorem{proposition}{Proposition}[section]
\newtheorem{definition}[proposition]{Definition}
\newtheorem{remark}[proposition]{Remark}

\begin{document}

\title{Ultraviolet Finite Quantum Field Theory on Quantum Spacetime}

\author{{D.\ Bahns\thanks{II. Institut f\"ur Theoretische Physik, 
Universit\"at Hamburg, Luruper Chaussee 149, D-22761 Hamburg. E-mail:
{\tt dorothea.bahns@desy.de, klaus.fredenhagen@desy.de~.}}}, 
{S.\ Doplicher\thanks{Dipartimento di Matematica, Universit\`a di Roma 
``La Sapienza'', P.le Aldo Moro~2, 00185 Roma. E-mail: {\tt dopliche@mat.uniroma1.it, piacitel@mat.uniroma1.it}~.  Research supported by MIUR and 
GNAMPA-INDAM.}}, 
K.\ Fredenhagen$^*$, G.\ Piacitelli$^\dagger$}

\renewcommand{\theequation}{\thesection.\arabic{equation}}
\numberwithin{equation}{section}

\date{January 15, 2003; revised March 20, 2003}

\maketitle

\centerline{\em Dedicated to Rudolf Haag on the occasion of his 
80${}^{\text{th}}$ birthday.}

\begin{abstract}
We discuss a formulation of quantum field theory on quantum space time where
the perturbation expansion of the $S$-matrix is term by term ultraviolet
finite.

The characteristic feature of our approach is a quantum version of the Wick
product at coinciding points: the differences of coordinates $q_j
- q_k$ are not set equal to zero, which would violate the
commutation relation between their components. We show that
the optimal degree of approximate coincidence 
can be defined 
by the
evaluation of a conditional expectation which replaces each
function of $q_j- q_k$ by its expectation value in optimally
localized states, while leaving the mean coordinates
$\tfrac{1}{n}(q_1 +\dotsb + q_n)$ 
invariant.

The resulting procedure is to a large extent unique, 
and is invariant under
translations and
rotations, but violates Lorentz invariance. Indeed, optimal
localization refers to a specific Lorentz frame, where the electric and
magnetic parts of the commutator of the coordinates have to 
coincide~\cite{dfr}.

Employing an adiabatic switching, we show that   the S-matrix is
term by term finite. 
The matrix elements of the transfer matrix are determined, at each order in
the perturbative   expansion, by kernels with Gaussian decay in the
Planck scale. 
The adiabatic limit and the 
large scale limit of this theory 
will be studied elsewhere.

\end{abstract}

\section{Introduction}

Spacetime quantization was proposed earlier than  
renormalization theory as a
possible way of regularizing quantum field theory \cite{snyder}.
Recently, a deeper motivation was given \cite{dfr_2,dfr}:  
the concurrence of the  
principles of quantum mechanics and of classical general relativity leads
to spacetime uncertainty relations;  the natural geometric
background that implements those relations is a noncommutative model
of spacetime. More precisely, in order to give localization in
spacetime  an operational meaning, 
{\em the energy transfer
associated to the localization of
an event by the Heisenberg uncertainty principle 
should be limited so that the generated gravitational field does
not trap the event itself  inside an horizon; otherwise 
the observation would be prevented.} 
This principle implies {spacetime uncertainty relations} which in a weaker form
can be written as
\begin{gather*}
\Delta q^0(\Delta q^1+\Delta q^2+\Delta q^3)\gtrsim \lambda_P^2,\\
\Delta q^1\Delta q^2+\Delta q^2\Delta q^3+\Delta q^3\Delta q^1\gtrsim
\lambda_P^2,
\end{gather*}
where $\lambda_P$ is the Planck length
$\sqrt{G\hbar c^{-3}}\simeq 1.6\times 10^{-33}\text{cm}$.
It is possible to implement exactly these relations by appropriate commutation
relations between the components of the spacetime coordinates $q^\mu$
\cite{dfr,dfr_2}
\begin{gather}
\label{q1}
[q^\mu,q^\nu]=i\lambda_P^2Q^{\mu\nu},\\
\label{q2}
[q^\mu,Q^{\nu\rho}]=0,\\
\label{q3}
Q_{\mu\nu}Q^{\mu\nu}=0,\\
\label{q4}
\left(\frac{1}{2}Q_{\mu\nu}(*Q)^{\mu\nu}\right)^2=I,
\end{gather}
where $*Q$ is the Hodge dual of $Q$.
These relations are covariant under the full Poincar\'e group. The irreducible
 representations of the spacetime commutation relations~(1) take the familiar
form (in absolute units, where $\lambda_P=1$)
\begin{equation}
\label{familiar}
[q^\mu,q^\nu]=i\sigma^{\mu\nu}I,
\end{equation}
where $\sigma$ is a real antisymmetric matrix
in the manifold $\Sigma$ defined by the conditions (\ref{q3},\ref{q4}) with
$Q^{\mu\nu}=\sigma^{\mu\nu}I$. They evidently break Lorentz covariance.
Interest in the relations (\ref{familiar}) was more recently raised by the
occurrence of closely related forms of noncommutativity
 also in string theory \cite{connes,schomerus}.
There exists, however,
an essentially unique, fully covariant representation where
the pairwise commuting, selfadjoint operators $Q^{\mu\nu}$ have the full
manifold $\Sigma$ as their joint spectrum.
The generalized Weyl correspondence
\[
\mathcal W(g\otimes f)= g(Q)f(q)
\]
extends to any symbol $F\in\mathcal C_0(\Sigma\times \mathbb R^4)$,
$\check F(\sigma,\cdot)\in L^1(\mathbb R^4)$, where $\check F(\sigma,\cdot)$ 
is the inverse Fourier transform of $F(\sigma,\cdot)$, for $\sigma$ fixed.
In the above
equation, $g(Q)$   
is to be understood in the sense of 
the joint functional calculus of the $Q^{\mu\nu}$s,
and
\[
f(q)=\int_{\mathbb R^4} dk\;\check f(k)e^{ikq},
\]
where $\check f(k)=(2\pi)^{-4}\int dx\;f(x)e^{-ik x}$ 
and   $kq=k_\mu q^\mu$, $kx=k_\mu x^\mu$.
The above correspondence induces a generalized twisted   product
\[
(F_1\star F_2)(\sigma,\cdot)=F_1(\sigma,\cdot)\star_\sigma F_2(\sigma,\cdot),
\]
on the symbols, by $\mathcal W(F_1\star F_2)=\mathcal
W(F_1)\mathcal W(F_2)$. Moreover, $\mathcal W(\overline
F)=\mathcal W(F)^*$. We denote by $\mathcal E$ the enveloping
C*-algebra of the resulting algebra; it is isomorphic
  to $\mathcal
C_0(\Sigma,\mathcal K)$, the C*-algebra of the continuous
functions taking values in the algebra $\mathcal K$ of the compact
operators on the
separable, infinite dimensional Hilbert space,
and vanishing at infinity.  The Poincar\'e group acts on the
symbols in $\mathcal E$ by
\[
(\tau_{(a,\varLambda)}F)(\sigma,x)=\text{det}\;\varLambda\;
F(\varLambda^{-1}\sigma{\varLambda^{-1}}^t,\varLambda^{-1}(x-a)).
\]
The manifold $\Sigma$ is the orbit of the standard symplectic matrix
$\sigma_0^{\mu\nu}\nolinebreak=\nolinebreak{
\mbox{\footnotesize\renewcommand{\arraystretch}{0.5} $ \left(
\begin{array}{rr}0&-\bf 1\\ \bf 1&0\end{array}\right)$}}\,,$ 
under the action $\sigma\mapsto\varLambda\sigma\varLambda^t$ of the
full Lorentz group.

By definition, the states of $\mathcal E$ with {\em optimal
localization} 
 (both in space and in time) minimize $\sum_\mu(\Delta
q_\mu)^2$.  This characterization is evidently invariant under
rotations and translations, but not under Lorentz boosts. It can
be shown (see \cite{dfr}) that the optimally localized states are
of the form 
\[
\langle\omega_a,F\rangle=
\int_{\Sigma_1}\mu(d\sigma)
(\eta_a F)(\sigma),
\quad F\in\mathcal E,
\]
where $\mu$ is any probability measure on the
distinguished subset $\Sigma_1$ of $\Sigma$, 
the orbit of $\sigma_0$ under the action of
$O(\mathbb R^3)$, and
$\eta_a:\mathcal E\rightarrow\mathcal C(\Sigma_1)$ is the localization
map with localization centre $a\in\mathbb R^4$,
\begin{equation}
\label{loc_map} (\eta_a F)(\sigma)= \int_{\mathbb R^4} dk\;\check %
F(\sigma,k) \exp\left\{-\frac {1}{2} \sum_{\mu=0}^3{k_\mu}^2
\right\}\,e^{ika}.
\end{equation}
In what follows, we will need only the localization map with localization 
centre $a=0$, and in order to simplify the notation we will denote it by 
$\eta$. However, the results below also hold for a general $a\in \mathbb R^4$.

It is convenient to introduce the enveloping
C*-algebra $\mathcal E_1$ generated by the restrictions $\gamma
F=F\restriction_{\Sigma_1}$ of the symbols to $\Sigma_1$. Then the
localization map $\eta$ is the composition $\eta=\eta^{(1)}\circ\gamma$
of the restriction map $\gamma:\mathcal E\rightarrow\mathcal E_1$,
with a positive map $\eta^{(1)}$ from $\mathcal E_1$ to $\mathcal
C(\Sigma_1)$, which is a conditional expectation in the sense
that\footnote{Note that, while $\mathcal C(\Sigma_1)$ is not a
subalgebra of ${\mathcal E}_1$, it is a subalgebra (actually,
the centre) of the multiplier algebra $M(\mathcal E_1)$.}
$\eta^{(1)}(zF)=z\eta^{(1)}(F)$, $z\in\mathcal C(\Sigma_1)$,
$F\in\mathcal E_1$.  $\eta$ will also denote  the normal extension to the 
multiplier algebra $M(\mathcal E)$. Then, 
\begin{equation}
\label{loc_car}
\big\langle\eta,e^{ikq}\big\rangle=e^{-\frac{1}{2}\sum_\mu{k_\mu}^2}\,,
\end{equation}
as a constant function of $\sigma\in\Sigma_1$.

By analogy with the
definition of $f(q)$, the evaluation of an ordinary quantum field $\phi$ on
the quantum spacetime is given  by  
\begin{equation*}
\phi(q)=\int_{\mathbb R^4} dk\; 
e^{ikq}\otimes\check\phi(k)
\end{equation*}
and is to be interpreted as a map from states on $\mathcal E$ to smeared
field operators,
\[
\omega\mapsto\phi(\omega)=\langle\omega\otimes\id,\;\phi(q)\rangle=
\int_{\mathbb R^4} dx\;\phi(x)\psi_\omega(x),
\]
where the r.h.s. is a   quantum field on the
ordinary spacetime, smeared with the test function $\psi_\omega$
defined by $\check\psi_{\omega}(k)=\langle\omega,\;
e^{ikq}\rangle$. 
If products of fields are evaluated in a state, the r.h.s. will
in general involve nonlocal expressions.


As in ordinary quantum field theory, due to
the singular properties of fields, products of fields are not {\em a priori}
well-defined. On the ordinary Minkowski spacetime, well-defined products of
fields are given by the so-called Wick products. They may be defined by
bringing the positive and negative frequency parts of the fields in the product
into ``normal order'', which in momentum space corresponds to putting all
destruction operators to the right. Another definition, which, contrary to
normal-ordering may also be applied on curved spacetimes, is given in terms of
the formal evaluation on the diagonal of a suitably subtracted product; one has,
for instance, at
second order $\wick{\phi(x)^2}=\lim_{x\rightarrow y} \, (\, \phi(x)\phi(y)-(
\Omega, \phi(x)\,\phi(y) \Omega) \,)$. 
The two constructions, while equivalent on the
ordinary Min\-kowski spacetime, lead to inequivalent generalizations
on the quantum spacetime.

In~\cite{dfr}, for instance, an interaction Lagrangean was given in terms of
the {\em usual} normal ordering of positive and negative frequency parts of
$\phi(q)^n$,
\begin{equation}
\label{old_wick}
\mathcal L_I(x)=\wick{(\phi\star\dotsb\star\phi)(x)}.
\end{equation}
Another possibility will be investigated in~\cite{bdfp2}. There, we consider
products of fields at different points as they arise in the context of the
Yang-Feldman equation, 
$\phi(q+x_1)\dotsb\phi(q+x_n)$, $x_i \in \mathbb R^4$. We then 
define the so-called quasiplanar Wick products by allowing only
terms which are local in a certain sense to be subtracted, and show that they
are well-defined on the diagonal, i.e. in the limit of coinciding points where
$x_i=x_j$. 

In this paper we consider yet another approach. The evaluation on the
diagonal is replaced by a suitable generalization compatible with the
uncertainty relations, leading to a regularized nonlocal effective interaction. 
The idea is that a product of fields at different
points, $\phi(q_1)\dotsm\phi(q_n)$, may be defined by
interpreting $q_1,\dotsc,q_n$ as {\em mutually independent quantum coordinates},
that is, by defining
\begin{equation}
\label{ind_coord}
{q_j}^\mu=I\otimes\dotsm\otimes I\otimes q^\mu\otimes I
\otimes \dotsm\otimes I\quad
\text{($n$ factors, $q^\mu$ in the \thh{j}\ slot)},
\end{equation}
and
\[
\phi(q_1)\dotsm\phi(q_n)=\int dk_1\dotsm dk_n\;\check\phi(k_1)
\dotsm\check\phi(k_n)e^{i(k_1q_1+\dotsb+k_nq_n)}.
\]
Now, the different spacetime components of each variable $q_j-q_k$, $j\neq k$,
no longer commute with one another, hence the limit $q_j-q_k\longrightarrow 0$
loses its natural meaning. 
We can, however, identify the central elements, i.e. take
$[q_{j}^\mu,q_{j}^\nu]=iQ^{\mu\nu}$ for all~$j$. This amounts to taking the 
tensor products in~(\ref{ind_coord}) not over the complex numbers, but over 
the centre 
$Z=\mathcal C_0(\Sigma)$ of (the multiplier algebra of) $\mathcal
E$. The limit
$q_j-q_k\longrightarrow 0$ will then be replaced by 
a {\em quantum diagonal map} 
which on each function of $q_j-q_k$ evaluates a state minimizing the
square Euclidean
length, while leaving the mean coordinates   invariant 
  (cf.~\cite{hesselberg},~\cite{cho}).

As a consequence of taking the tensor product in~(\ref{ind_coord}) over~$Z$, 
the mean coordinates $\tfrac{1}{n}\sum_jq_j$ commute 
with the relative coordinates $q_j-q_k$ (in the strong sense), e.g. for $n=2$, 
\begin{equation}
\label{tensor}
\left[(q_1+q_2)^\mu,(q_1-q_2)^\nu\right]=
[q^\mu,q^\nu]\otimes I-
I\otimes[q^\mu,q^\nu]= 0.
\end{equation}
This fact turns out to be crucial for the construction of the quantum
diagonal map and provides an additional motivation for taking the
tensor product over $Z$,

Actually, the tensor product $\mathcal E\otimes_Z\mathcal E$ of $Z$-moduli
can be defined as the completion relative to the maximal C*-seminorm of
the quotient of the algebraic tensor product over $\mathbb C$ modulo
the two sided ideal generated by the multiples in $\mathcal E\odot\mathcal E$
of $I\odot z-z\odot I$, where $z$ varies in $Z$. It may be equivalently
described as
the fibrewise tensor product of bundles of C*-algebras (continuous
fields of C*-algebras, in the terminology of \cite{dixmier}) over
$\Sigma$.

Another motivation to use a $Z$-module tensor product is to view the components
$Q^{\mu\nu}$ of the coordinates' commutator as universal data which are
the same for the different variables corresponding to independent events. 
The $Q^{\mu\nu}$s are thus treated as a point independent geometric 
background, which, however, is translation invariant and Lorentz covariant.

Since the C*-algebra $\mathcal E$ describes the regular representations of
(1), i.e. integrable to
a representation of the Weyl relations  
\[
e^{ik q}e^{ih q}=e^{-\frac{i}{2}k_\mu h_\nu Q^{\mu\nu} }
e^{i(k+h) q},
\]
the uniqueness theorem of von Neumann~\cite{neumann}, 
applied to each fibre at $\sigma\in\Sigma$, ensures that
commutativity implies tensor factorization over $Z$. This fact will allow us
to obtain the desired map as a conditional expectation.

Furthermore,   we will use the tensor product of $n+1$ copies of
the basic algebra as an auxiliary algebra, where the mean coordinates are
(affiliated to the algebra) in the first factor, and where 
the algebra to which the
difference variables $q_j-q_k$ are affiliated is identified with a
subalgebra of the auxiliary algebra, associated to the factors from
  slot
$2$ to $n+1$. The desired quantum diagonal map
\[
E^{(n)}:\mathcal E\otimes_Z\dotsm\otimes_Z\mathcal E\longrightarrow\mathcal E_1
\]
is then obtained by   evaluating
$\gamma\otimes_Z\eta^{n\otimes_Z}$ on such tensor products, 
where $\eta$ is the localization
map~(\ref{loc_map}) with localization centre $a=0$, and where 
$\gamma:\mathcal E\rightarrow\mathcal E_1$
is the restriction map. It  turns out that the application of
$E^{(n)}$ to functions whose symbol do not depend on $Q$
explicitly, yields expressions which in turn are independent of $Q$.

The quantum diagonal map replaces the ordinary evaluation
at coinciding points. Contrary to the ordinary case, it
yields a well-defined expression when applied to a product of fields, 
\begin{eqnarray*}
\phi^{(n)}(q)&=&E^{(n)}(\phi(q_1)\dotsm\phi(q_n))
\\&=&
\int d^4k_1\dotsm d^4k_n\; r_n(k_1,\dotsc,k_n)\,
\check\phi(k_1)\dotsm\check\phi(k_n)\,e^{i(k_1+\dotsb+k_n){q}}
\end{eqnarray*}
since a nonlocal regularizing kernel $r_n$ appears. We conclude that contrary 
to the ordinary case, no infinite counterterms have to be subtracted and 
$\phi^{(n)}(q)$ may be used directly to define the interaction in the quantum
theory. 
Regarding the combinatorics, it is, however, convenient to 
additionally apply ordinary normal ordering, and to define a
{\em quantum Wick power} as
\begin{eqnarray*}
\wick{\phi^n(q)}_Q&=&E^{(n)}(\wick{\phi(q_1)\dotsm\phi(q_n)})
\\&=&
\int d^4k_1\dotsm d^4k_n\; r_n(k_1,\dotsc,k_n)\,
\wick{\check\phi(k_1)\dotsm\check\phi(k_n)}\,e^{i(k_1+\dotsb+k_n){q}}\,.
\end{eqnarray*}
The quantum Wick power $\wick{\phi(q)^n}_Q$ as well as $\phi^{(n)}(q)$ 
may be understood as functions of
$q$, not explicitly depending on $Q$, taking values in the field
operators. In other words, $\wick{\phi(q)^n}_Q$ and $\phi^{(n)}(q)$ 
formally are elements
of $\mathcal E_1\otimes\mathfrak F$, where $\mathfrak F$ is the field algebra.

Once products of fields are given a precise meaning, one may apply an
appropriate perturbative setup. Since sharp localization in time is compatible
with the spacetime uncertainty relations (at the cost of complete
delocalization in space), {\em one} possibility is, for instance, to follow the
standard approach to perturbation theory in the interaction representation,
involving integrations at sharp fixed times~\cite{dyson}. If the Lagrangean is
symmetric, the resulting $S$-matrix is formally unitary by construction (at
least before renormalization). 

In~\cite{dfr}, such an approach was proposed, based on the interaction
Lagrangean~(\ref{old_wick}). Unfortunately, the resulting perturbation theory
is not free of ultraviolet divergences. 
This fact was first observed in \cite{filk} where, 
however, instead of the interaction picture used in 
\cite{dfr}, the theory was defined in terms of modified 
Feynman rules which may be formally derived from a path integral 
formulation. As first observed in \cite{unitarity} the resulting 
theory violates unitarity, a defect which may be traced 
back to the problem of time ordering on a 
(space/time)-noncommutative theory, as discussed in 
\cite{bdfp1}: the time ordering naturally defined in the interaction
picture formulation (cf \cite[eq. (6.15)]{dfr} and subsequent comments, as
well as section \ref{diagramm} of the present paper) 
does not violate unitarity.
 As a consequence, the formulation of the 
theory in terms of modified Feynman rules is not equivalent 
to the one discussed here.

Another inequivalent approach, which, however, yields a 
unitary perturbation theory was proposed in \cite{bdfp1}. 
This approach is based on the Yang-Feldman equation and 
will be discussed elsewhere \cite{bdfp2}. 

Instead, we will again apply the standard approach to perturbation theory in the
interaction representation, this time employing the quantum Wick products. 
 
The interaction Hamiltonian on the quantum spacetime is then given by
\[
\mathcal H_I(t)=\lambda\int_{q^0=t}d^3q\;\wick{\phi(q)^n}_Q
\]
as a constant operator--valued function  of 
$\Sigma_1$ (i.e. $\mathcal H_I(t)$ is formally in
$\mathcal C(\Sigma_1)\otimes\mathfrak F$).  
While in~\cite{dfr} one still had to handle the dependence of the Hamiltonian on
$\sigma$, in  the approach adopted here, $\mathcal H_I(t)$ is a {\em constant}
function of $\sigma\in \Sigma_1$. As a consequence, our procedure leads to a 
unique prescription for the interaction Hamiltonian on quantum spacetime.

The resulting effective non local Hamiltonian is
\begin{equation}
\label{H_I}
H_I(t)=\lambda\int d^3\bx\;\Leff(t,\bx)
\end{equation}
where $\Leff$ is the effective nonlocal interaction Lagrangean
\begin{gather}
\nonumber
\Leff(x)= c_n\int_{\mathbb R^{4n}} da_1\dotsm
da_n\;\wick{\phi(x+a_1)\dotsm\phi(x+a_n)}\\
\label{L_eff}
\exp\bigg\{-\frac{1}{2}\sum_{j,\mu}{a_j^\mu}^2\bigg\}
\delta^{(4)}\bigg(\frac{1}{n}\sum_{j=1}^n{a_j}\bigg).
\end{gather}

It will be shown in section \ref{finite} that the corresponding
perturbation theory
is free of ultraviolet divergences.
The ultraviolet regularization arises as a point--split regularization by
convolution with Gaussian kernels,  
and we will show that, by insertion of an adiabatic
switch, the perturbation series is order by order finite, and each term is
a well defined, closed operator with a common core. 
The only remaining source of divergences
  is then given by possible infinite volume effects arising in the 
adiabatic limit, which will be discussed
 elsewhere. 

The ultraviolet finiteness of the theory presented in this 
paper is in accordance with the expectation that 
noncommutativity of spacetime may regularize the theory. 
Other examples for ultraviolet finite theories on 
noncommutative spaces were discussed in \cite{CDP98}, for 
instance compact spacetimes, corresponding to finite dimensional 
algebras.

It is noteworthy that the transition matrix elements
will vanish as Gaussian functions of the energies and momentum transfers
expressed in Planck units.

While in the high energy limit the transition amplitudes vanish rapidly as
a result of the quantum delocalization of the interaction, in the low
energy limit one would expect that the corrections to the ordinary theory
on Minkowski space vanish. This is clearly possible only after a finite
renormalisation; the structure of the needed counterterms and the
dependence upon the Planck length of the renormalisation constants will be
studied elsewhere.  

Note that in the limit where the Planck length
can be neglected, the renormalized theory on quantum spacetime should
coincide with the ordinary renormalized theory on Minkowski space. 
  At the physical values of the Planck length, the
effect of the quantum nature of spacetime should manifest itself as quadratic
or higher order corrections, since gravitation is not explicitly taken into
account, but manifests itself only through the commutator of the
coordinates.

A weak point of the approach to quantum field theory on quantum
spacetime presented here is that, while, as was   first shown in \cite{dfr}, the
prescription leading to (\ref{old_wick}) does not alter the {\em free} 
Hamiltonian, the
prescription discussed here would indeed change it, 
replacing it by a deformed operator which
would no longer be the zero component of a Lorentz vector. 
We therefore treat, in this paper, the interaction on a 
different footing than the unperturbed Hamiltonian which we 
identify with that of the usual free theory\footnote{There exists 
an alternative approach, based on the action principle, 
which avoids this unsatisfactory feature. It will be 
discussed in a forthcoming publication \cite{bdfp3}.}.
As a consequence, Lorentz invariance is violated in an essential way,
since
optimally localized states are defined relative to a particular Lorentz frame.
However, spacetime translation and space rotation invariance are preserved.
Moreover, the evaluation of optimally localized states on the difference
variables $q_j-q_k$ automatically restricts the joint eigenvalues
$\sigma^{\mu\nu}$ of $Q^{\mu\nu}$ to $\Sigma_1$, the basis of $\Sigma$,
where the electric and magnetic parts of $\sigma$ are equal or opposite.
This gives an a posteriori motivation for a similar choice, made in \cite{dfr},
which was motivated by simplicity and by the need of preserving space rotation
invariance.

\section{The quantum diagonal map}
\label{main}

According to the previous discussion,
\[
{q_j}^\mu=\underbrace{I\otimes_Z\dotsm\otimes_Z I\otimes_Z q^\mu\otimes_Z I
\otimes_Z \dotsm\otimes_Z I}_{\text{$n$ factors}},\quad
\text{$q^\mu$ in the \thh{j}\ slot},
\]
fulfill the relations (for any $i,j=1,\cdots,n$)
\begin{gather}
\label{qn1}
[{q_i}^\mu,{q_j}^\nu]=i\lambda_P^2\delta_{ij}
{\mathcal Q}^{\mu\nu},\\
\label{qn2}
[{q_j}^\mu,{\mathcal Q}^{\nu\rho}]=0,\\
\label{qn3}
{\mathcal Q}_{\mu\nu}{\mathcal Q}^{\mu\nu}=0,\\
\label{qn4}
\left(\frac{1}{2}{\mathcal Q}_{\mu\nu}(*\mathcal Q)%
^{\mu\nu}\right)^2=I.
\end{gather}
The correspondence
\[
\mathcal W^{(n)}
(g\otimes f)=g(\mathcal Q)f(q_1,\dotsc,q_n),\quad
g\in\mathcal C_0(\Sigma),
f\in\mathcal C_0(\mathbb R^{4n}),\check f\in L^1(\mathbb R^{4n}),
\]
extends to the generalized symbols $F=F(\sigma,x_1,\dotsc,x_n)$
as   usually, where
\[
f(q_1,\dotsc,q_n)=\int dk_1\dotsc,k_n\;\check f(k_1,\dotsc,k_n)
e^{i(k_1q_1+\dotsb+k_nq_n)}.
\]
It induces a product and an involution on
the generalized $n$-symbols, and the enveloping C*-algebra of the
resulting algebra is precisely $\mathcal E^{(n)}=\mathcal
E\otimes_Z\dotsm\otimes_Z\mathcal E$.

\begin{remark}
\label{ideals}
Note that, since $\mathcal K\otimes\mathcal K\sim\mathcal K$ as C*-algebras,
$\mathcal E\otimes_Z\dotsm\otimes_Z\mathcal E\sim\mathcal E\sim
\mathcal C(\Sigma,\mathcal K)$. Closed 2-sided ideals $J$ in 
$\mathcal E \otimes_Z\dotsm\otimes_Z\mathcal E$ are then in a 1-1 
correspondence with closed ideals in $Z$ (the kernel of the restriction to $Z$
of the canonical extension to $M(\mathcal E)$ of the projection map
mod $J$), hence are in a 1-1 correspondence with the closed subsets of
$\Sigma$.
\end{remark}

Let us now introduce the coordinates of the mean event, denoted 
mean coordinates for short,
$$
{\bar q}=\frac{1}{n}({q_1}+\dotsb+{q_n})
$$
as well as the separations
$$
q_{ij} =q_{i}-q_j.
$$
Then
\begin{equation}
\label{fact}
q_i={\bar q}+\frac{1}{n}\sum_jq_{ij}.
\end{equation}
Since the commutator
$[{q_j}^\mu,{q_j}^\nu]=i{\mathcal Q}^{\mu\nu}$
does not depend on $j$, the following
strong commutation relations hold:
\begin{eqnarray}
\label{qbar_CR}
e^{ik_\mu{\bar q}^\mu}e^{ik'_\mu{\bar q}^\mu}&=
&e^{-i\frac{1}{2n}(\mathcal Q^{\mu\nu}k_\mu k'_\nu)}
e^{i(k+k')_\mu{\bar q}^\mu},\\
\label{qbar_CR2}
e^{ik_\mu{\bar q}^\mu}e^{ik'_\mu q_{ij}^\mu}&=
&e^{ik'_\mu q_{ij}^\mu}e^{ik_\mu{\bar q}^\mu}.
\end{eqnarray}

We have the following factorization. Let $\tilde q$ be
coordinates with characteristic length $1/\sqrt n$, i.e. $[\tilde q^\mu,\tilde
q^\nu]=\,\frac{i}{n}\,Q^{\mu\nu}$. Define
\begin{equation}
\label{first}
\bar\bq^\mu:=\tilde q^\mu\otimes_ZI^{n\otimes_Z},\quad
\quad {\bq_{ij}}^\mu=I\otimes_Z {q_{ij}}^\mu,
\end{equation}
and
\begin{equation}\label{bqi}
{\bq}_i:=\bar\bq+\frac{1}{n}\sum_j
\bq_{ij}.
\end{equation}
We immediately check that the above elements 
also fulfill the relations  (\ref{qn1}--\ref{qn4}) in the regular
form, where $[{\bq_j}^\mu,{\bq_j}^\nu]=iQ^{\mu\nu}$.
By  von Neumann uniqueness (at each fixed $\sigma$; see
\cite{dfr}), there exists a faithful *-homomorphism
\[
\beta^{(n)}:\mathcal E{(n)}\mapsto M(\mathcal E^{(n+1)})
\]
such that
\[
\beta^{(n)}(q_i)=\bq_i.
\]
This follows from the fact that regularity implies that the map 
$q^{\mu}_i\mapsto\bq^\mu_i$ determines  
a *-homomorphism $\beta_i:\mathcal E\rightarrow M(\mathcal E^{(n+1)})$ (whose
canonical extension to $M(\mathcal E)$ will still be denoted by $\beta_i$);
the ranges of $\beta_i$ and $\beta_j$ commute for $i\neq j$ and 
$\beta_i\restriction_Z$ is an isomorphism independent of $i$. By the universal 
properties of the tensor product and its uniqueness for nuclear C*-algebras
(as $\mathcal E$), there is a *-homomorphism $\beta^{(n)}$ of 
$\mathcal E^{(n)}$ to $M(\mathcal E^{(n+1)})$, s.t.
$\beta(A_1\otimes_Z\dotsm\otimes_Z A_n)=\beta_1(A_1)\dotsm\beta_n(A_n)$, 
$A_j\in\mathcal E$.
By assumption, $\beta^{(n)}$ is faithful on $Z$, hence, by remark \ref{ideals} 
on page \pageref{ideals}, $\beta^{(n)}$ is faithful.

Explicitly, 
\[
\beta^{(n)}\big(g(\mathcal Q)f({q}_1,\dotsc,{q}_n)\big)
=g(Q)f({\bq}_1,\dotsc,{\bq}_n),
\]
where, of course,
\[
f({\bq}_1,\dotsc,{\bq}_n)=\int dk_1\dotsm dk_n\;\check
f(k_1,\dotsc,k_n)e^{ik_1{\bq}_1}\dotsm e^{ik_n{\bq}_n}.
\]

\begin{definition} The quantum diagonal map
$E^{(n)}:\mathcal E^{(n)}\rightarrow \mathcal E_1$ is defined as
\[
E^{(n)}=\big(\gamma\otimes_Z\underbrace{\eta\otimes_Z\dotsm\otimes_Z\eta}%
_{\text{$n$ factors}}\big)\circ\beta^{(n)},
\]
where $\eta,\gamma$ are the localization map and the restriction to
$\Sigma_1$ (projection of $\mathcal E$ onto $\mathcal E_1$),
respectively. Note that the generators $\tilde q^\mu$ of the algebra
in which $E^{(n)}$ takes values have characteristic length $1/\sqrt n$.
\end{definition}

To motivate this choice, let us 
recall that the difference variables 
${q_{ij}}^{\mu}/\sqrt 2$ 
fulfill the commutation relations (\ref{q1}--\ref{q4}), and a short computation yields
\[
\bigg\langle\underbrace{\eta\otimes_Z\dotsm\otimes_Z\eta}_{\text{$n$ factors}},
e^{ik_\mu {q_{ij}}^\mu/\sqrt 2}\bigg\rangle=
\big\langle\eta,e^{ik_\mu q^\mu}\big\rangle
\]
(as constant functions of $\sigma\in\Sigma_1$; compare with (\ref{loc_car})).
In other words, $\eta^{n\otimes_Z}$ minimizes the Euclidean separation 
$\sum_\mu({q_i}^\mu-{q_j}^\mu)^2$. $E^{(n)}$ will also
denote the normal extension
of the above map to the multiplier algebra $M(\mathcal E^{(n)})$.

Note also that, had we used $\eta_a$ instead of $\eta=\eta_0$, we would
have defined the same map $E^{(n)}$, since the separations $q_i-q_j$
are invariant under translations.
\begin{proposition}
Let $f\in\mathcal C_0(\mathbb R^{4n})$, $\check f\in L^1(\mathbb R^{4n})$.
The explicit form of the quantum diagonal map on $f$ is given by
\[
E^{(n)}(f(q_1,\dotsc, q_n))=
\int_{\mathbb R^{4n}} dk_1\dotsm dk_n \check f(k_1,\dotsc,k_n)
    r_n(k_1,\dotsc,k_n)
    e^{i\left(\sum_ik_i\right)\tilde q}\nonumber
\]
where 
\begin{equation}
\label{rn}
r_n(k_1,\dotsc,k_n)=\exp\left\{-\frac{1}{2}
\sum_{\mu=0}^3\left(\sum_{j=1}^n{{k_j}_\mu}^2-
\frac{1}{n}\sum_{j,l=1}^n{k_j}_\mu {k_l}_\mu\right)\right\}.
\end{equation}
Equivalently,
\[
E^{(n)}\big(f(q_1,\dotsc,q_n)\big)=h(\tilde q)
\]
where
\[
h(x)=c_n\int_{\mathbb R^{4n}} da_1\dotsm da_n\;f(x+a_1,\dotsc,x+a_n)
\hat r_n(a_1, \dotsc,a_n),
\]
with $c_n=n^2(2\pi)^{-8(n-1)}$ 
and, with $|a|^2=\sum_{\mu=0}^3 a_\mu a_\mu$,
\[
\hat r_n(a_1, \dotsc,a_n)
=\exp\big(\,{-\frac{1}{2} |a_1|^2 -\dots -\frac{1}{2} |a_n|^2}\,\big)\,
\delta^{(4)}\big(\,\frac{1}{n}\sum_{j=1}^n a_j\,\big).
\]
In particular, $E^{(n)}\big(f(q_1,\dotsc,q_n)\big)$ is a constant function of
$\sigma\in\Sigma_1$.
\end{proposition}
\noindent{\em Proof.} A simple computation yields, by the the definition
(\ref{bqi}) of ${\bq}_i$,
\[
\exp\left\{i\sum_jk_j{\bq}_j\right\}=
\exp\left\{i\bigg(\sum_jk_j\bigg)\bar q\right\}\otimes_Z
\exp\left\{i\sum_j\bigg(k_j-\frac{1}{n}\sum_lk_l\bigg)q_j\right\}.
\]
By the above and (\ref{qbar_CR}),(\ref{qbar_CR2}), we have
\begin{eqnarray*}
\lefteqn{
\left\langle\gamma\otimes_Z\eta^{n\otimes_Z},\;
f({\bq}_1,\dotsc,{\bq}_1)\right\rangle=}\\
&=&\int dk_1\dotsm dk_n \check f(k_1,\dotsc,k_n)
\exp\left\{i
\left(\sum_jk_j\right)\tilde q\right\}\\
&&\prod_{j=1}^n\left\langle\eta,\;
\exp\left\{i\left(k_j-\frac{1}{n}\sum_{l}k_l\right){q}_j\right\}
\right\rangle
\end{eqnarray*}
as a constant function of $\sigma$; (\ref{rn}) 
then follows by a straightforward computation.
Standard computations provide the configuration space kernel.\qed

The quantum diagonal map takes a particular simple form if 
evaluated in optimally localized states. Indeed, let 
$\tilde{\eta}_{a}$ denote the localization map $\eta_{a}$ 
applied to the mean position coordinates $\tilde{q}$. Then 
a simple calculation yields the formula
\[\tilde{\eta}_{a}\circ E^{(n)}=\eta_{a}^{\otimes_{Z}^n} \ . \]
Since the function $a\to\eta_{a}(f)$ may be understood as 
the best commutative analogue of an element $f$ of the 
noncommutative algebra, this formula provides an additional 
justification of the present approach. (Cf. also the 
discussion in \cite{dfr} and \cite{CDP98}.)

\section{A class of ultraviolet finite theories on the quantum spacetime}
\label{finite}

The uncertainty relations (2) are compatible with sharp localization
in time, at the cost of total delocalization in space. Consistently,
the centre valued map
\[
g(Q)f(q)\mapsto g(Q)\int d^3\bx\;f(t,\bx)
\]
extends to a a positive partial trace $\int_{q^0=t}d^3q$ (see
\cite{dfr} for details), which commutes with the restriction $\gamma$ 
to $\Sigma_1$.

For a fixed choice of a frame of
reference, we formulate a traditional perturbative setup in the spirit of
\cite{dyson,bogo}. Consider for simplicity the $\lambda \phi^n$ interaction; then the
formal interaction Hamiltonian will be defined as 
\[
\mathcal H_I(t)=\lambda\int_{q^0=t}d^3{q}\;
\mathcal L_I(q),
\]
where the interaction Lagrangean $\mathcal L_I(q)$ 
may be either the \thh{n}\ {\em quantum Wick power}, 
defined by evaluating the quantum  
diagonal map on a normally ordered product of fields\footnote{We recall that a
monomial $A=a^\sharp(\psi_1)\dotsm a^\sharp(\psi_n)$ 
in  the creation an destruction operators ($a^\sharp=a,a^\dagger$) 
is called normally ordered or Wick ordered and denoted $\wick{A}$, 
if all creation operators stand left 
of the destruction operators.},
\begin{eqnarray*}
\wick{\phi^n(q)}_Q&=&E^{(n)}(\wick{\phi(q_1)\dotsm\phi(q_n)})
\\&=&
\int d^4k_1\dotsm d^4k_n\; r_n(k_1,\dotsc,k_n)\,
\wick{\check\phi(k_1)\dotsm\check\phi(k_n)}\,e^{i(k_1+\dotsb+k_n){q}}\,,
\end{eqnarray*}
or the {\em regularized product}, defined by evaluating the quantum
diagonal map on a product of fields as it stands, without application of normal
order, 
\begin{eqnarray*}
\lefteqn{\phi^{(n)}(q)\quad=\quad E^{(n)}(\phi(q_1)\dotsm\phi(q_n))}
\\&=&
\int d^4k_1\dotsm d^4k_n\; r_n(k_1,\dotsc,k_n)\,
\,\check\phi(k_1)\dotsm\check\phi(k_n)\,e^{i(k_1+\dotsb+k_n){q}}
\\&=&\int dk e^{ik{q}} 
\int dy e^{-iky}\,
\int d^4x_1\dotsm d^4x_n\; \hat r_n(y-x_1,\dotsc,y-x_n)
\,\phi(x_1)\dotsm \phi(x_n)\,.
\end{eqnarray*}
Clearly, the first definition yields a well-defined expression, but, contrary to
the ordinary case, normal ordering is not necessary due to the regulating kernel
$r_n$ which renders the second product well-defined as well\footnote{That this 
is true may be either checked by expanding the product of fields in normally
ordered products and check that all integrals are finite, or by employing the
method of wavefront sets to show that ${\rm Diag}\circ
(\hat r_n \times (\phi \dots \phi))$, where $\times$ denotes the ordinary
convolution, is well-defined.}. 
In fact, both the quantum 
Wick power $\wick{\phi(q)^n}_Q$ as well as $\phi^{(n)}(q)$ 
may be understood as functions of
$q$,   not explicitly depending on $Q$, 
taking values in the field
operators. In other words, $\wick{\phi(q)^n}_Q$ and $\phi^{(n)}(q)$ 
formally are elements of $\mathcal E_1\otimes\mathfrak
F$, where $\mathfrak F$ is the field algebra\footnote{In more rigorous 
mathematical terms,
$\wick{\phi(q)^n}_Q$ and $\phi^{(n)}(q)$ are affine maps from states 
on~$\mathcal E$ to quadratic forms.}.

In the following, however, we will  base our investigation on an interaction
given by a quantum Wick power.  For one thing, it simplifies the
combinatorics, and in view of the adiabatic limit, which we hope to study in a
later paper, normal ordering  may even be necessary.

The resulting Hamiltonian $\mathcal H_I(t)$ is formally affiliated to 
$\mathcal C(\Sigma_1)\otimes\mathfrak F$,
where $\mathfrak F$ is the free Bose field algebra on the ordinary spacetime. 
Roughly speaking, $\mathcal H_I(t)$ is a function from
$\Sigma_1$ to (formal) field operators, i.e. to quadratic forms.

In order to retrieve  the Hamiltonian of the equivalent theory on
the ordinary spacetime, one has to integrate over some probability measure
$\mu$ over $\Sigma_1$, defining $H_I(t)=\int d\mu\;\mathcal H_I(t)$. Since,
however, $\mathcal H_I(t)$ is a constant function of $\sigma$, the choice of
$\mu$ is irrelevant. This fact should be contrasted with the case  considered in
\cite{dfr}, where the non-irrelevant choice of a particular  measure~---~though
the most reasonable in that context~---~was to some extent  arbitrary, as well
as the special r\^ole played by $\Sigma_1$.

The resulting Hamiltonian $H_I(t)$
for the equivalent theory on the ordinary spacetime
can then be put in the form (\ref{H_I}), where the effective nonlocal 
Lagrangean is given by
\begin{gather*}
\Leff(x)=\int dk\;\check \Leff(k)e^{ikx}\,,
\\
\check{\mathcal L}_{\text{eff}}(k)=
\int d^4k_1\dotsm d^4k_n\; r_n(k_1,\dotsc,k_n)
\wick{\check\phi(k_1)\dotsm\check\phi(k_n)}\,\delta^{(4)}\bigg(k-
\sum_{j=1}^nk_j\bigg).
\end{gather*}

Note that in the perturbation series ((\ref{bogodyson}) here below) 
the time ordering of
products $H_I(t_1)\dotsm H_I(t_N)$ will refer to the variables $t_1,\dotsc,t_N$ rather than to the integration variables in (\ref{L_eff}).

The fundamental result of this section is that the finite 4-volume theory
yields a finite perturbation series. More precisely, we turn the coupling 
constant $\lambda$ into a smooth function of $x$ vanishing at infinity
sufficiently fast, of the form
$\lambda(t,\bx)=\lambda'(t)\lambda''(\bx)$, and we show that the 
corresponding Dyson series is well defined at all orders. Well-known methods   from ordinary quantum field theory are employed.

\begin{proposition}
For any Schwartz function $\lambda$ of the form
$\lambda(t,\bx)=\lambda'(t)\lambda''(\bx)$, 
$\lambda'\in\mathcal S(\mathbb R)$,
$\lambda''\in\mathcal S(\mathbb R^3)$, 
the formal series 
\begin{equation}
\label{bogodyson}
S[\lambda]=
T\exp\left\{-i\int d^4x\;\lambda(x)\Leff(x)\right\}=
I+\sum_{N=1}^\infty(-i)^N S^{(N)}[\lambda]
\end{equation}
is finite at all orders. More precisely, it is possible~---~
by Wick reduction~---~to put the \thh{N}\ order contribution
in the form of a finite sum of closable
operators with common core $D_{\mathcal S}$ (the subspace of the Fock space
consisting of the vectors with finitely many particles
and with Schwartz $n$-particle components for each $n$).
  By construction, $ S[\lambda]$ is unitary.
\end{proposition}

\begin{remark} While the existence of the adiabatic limit 
$\lambda\rightarrow 1$ is
questionable due to the breakdown of Lorentz covariance, 
the infinite volume limit
$\lambda''\rightarrow 1$ (with $\lambda'$ fixed) of the Gell-Mann--Low formula
for 
$S[\lambda'\otimes \lambda'']/\langle S[\lambda'\otimes\lambda'']\rangle_0$ 
exists as a
quadratic form. Indeed, the only terms in the perturbation expansion of
$S[\lambda'\otimes
\lambda'']$ which are divergent in the limit $\lambda''\rightarrow 1$ 
(with $\lambda'$
fixed) are precisely those containing vacuum--vacuum parts. The actual
behaviour of the full adiabatic limit will be investigated elsewhere.
\end{remark}

\noindent{\em Proof.}
We shall
follow standard conventions (see e.g.
\cite{reed_simon2}): in particular,
$a(g),a^{\dagger}(g)$ are the destruction and creation operators
on the symmetric Fock space
$\exp(L^{2}(\mathbb R^3))$, $g\in L^2(\mathbb R^3)$, and
\begin{gather}
\nonumber
a(g)=\int_{\mathbb R^3} d\bk\; a(\bk)\overline{g(-\bk)},\quad\quad
a^{\dagger}(g)=\int_{\mathbb R^3} d\bk\; a^\dagger(\bk)g(\bk),\\
\label{freefield}
\phi(t,\bx)=\frac{1}{(2\pi)^{3/2}}\int_{\mathbb R^3}
\frac{d\bk}{\sqrt{2\omega(\bk)}}\left\{
e^{ik_\mu x^\mu}a^\dagger(\bk)+
e^{-ik_\mu x^\mu}a(\bk)
\right\}
\end{gather}
as quadratic forms on $D_{\mathcal S}\times D_{\mathcal S}$, where
$\omega(\bk)=\sqrt{|\bk|^2+m^2}$, and $k=(\omega(\bk),\bk)$.

The cutoff Lagrangian $H_I^{(\lambda)}(t)$ is given by 
\[
H_I^{(\lambda)}(t)=\int_{\mathbb R^3}d\bx\;\lambda(t,\bx)\Leff(t,\bx)=
\lambda'(t)\int_{\mathbb R^3}d\bx\;\lambda''(\bx)\Leff(t,\bx).
\]
We introduce the following compact notations
\begin{gather}
\label{Smart1}
\underline a^{\mu}=(a_1^\mu,\dots,a_n^\mu)\in\mathbb R^{n},\\
\label{Smart2}
\underline{\boldsymbol a}=
({\boldsymbol a}_1,\dots,{\boldsymbol a}_n)\in\mathbb R^{3n},\\
\label{Smart3}
\underline a=(\underline a^0,\underline{\boldsymbol a})=
(a_1,\dots, a_n)\in\mathbb R^{4n}.\\
\intertext{The translation of all the $4$-vectors in $\ua$ by the same
$4$ vector $x$ will be denoted by}
\label{smart4}
\ua-x=(a_1-x,a_2-x,\cdots,a_n-x),\quad x,a_j\in\mathbb R^4.\\
\intertext{The symbol $\cdot$ will denote the canonical 
Euclidean scalar product in $\mathbb R^n$, $\mathbb R^{3n}$, 
$\mathbb R^{4n}$, depending on the context; then}
\label{Smart5}
\underline a\;\underline b=\underline a_{\mu}\cdot\underline b^\mu
=\ua^0\cdot\ub^0-\boldsymbol\ua\cdot\boldsymbol\ub,\\
\label{Smart6}
|\underline a|^2=\ua\cdot\ua.
\intertext{Moreover,}
\label{Smart7}
d\underline a=d\ua^0d\boldsymbol\ua=\prod_{\mu=0}^3d\underline a^{\mu}=
\prod_{\mu=0}^3\prod_{j=1}^n da^\mu_j.\\
\intertext{Finally, for any function $g=g(x)$ of $\mathbb R^4$, we write}
\label{Smart8}
g^{(n)}(\ux)=g(x_1)\dotsm g(x_n),\\
\intertext{and, in particular,}
\label{Smart9}
\wick{\phi^{(n)}(\underline x)}=\wick{\phi(x_1)\dots\phi(x_n)}.
\end{gather}
Standard computations yield 
\begin{align*}
S_N[\lambda]=&
\int_{{\mathbb R}^{4nN}}d\ua_1\dotsm d\ua_N\;
\bigg(\prod_{1\leq M< N}\theta\big(\kappa^0(\ua_{M+1}^0-\ua_M^0)\big)\bigg)\\
&G_\lambda(\ua_1)\dotsm G_\lambda(\ua_N)\;
\wick{\phi^{(n)}(\ua_1)}\dotsm\wick{\phi^{(n)}(\ua_N)},
\end{align*}
where $\kappa^0(\ua^0)$ is the time component of the mean point
\begin{equation}
\label{bary}
\kappa(\ua)=(\kappa^0(\ua^0),\boldsymbol\kappa(\boldsymbol\ua))
=\frac{1}{n}\sum_{j=1}^na_j\in\mathbb R^4
\end{equation}
of $\ua=(a_1,\dotsc,a_n)$, and 
\begin{equation}
\label{GSchwartz}
G_\lambda(\ua)=
\lambda(\kappa(\ua))e^{-\frac{1}{2}
|\ua-\kappa(\ua)|^2}
\end{equation}
is a Schwartz function.

With $P(I_n)$ the set of all subsets
of $I_n=\{1,\dotsm,n\}$
  (including the trivial subsets), we shall write
$J_\bullet$ for any choice of $N$ elements of $P(I_n)$, namely
$J_\bullet=\{J_1,\dotsc,J_N\}$ with $J_M\in P(I_n)$, $M=1,\cdots,N$.
Then, by (\ref{freefield}), we get 
\begin{align}
\nonumber
S_N[\lambda]=&\sum_{J_\bullet}
\int_{\mathbb R^{3nN}}d\ubk_1\dotsm
d\ubk_N
\frac{K_\lambda^{J_\bullet}(\ubk_1,\dotsc,\ubk_N)}
{\sqrt{\omega^{(n)}(\ubk_1)\dotsm \omega^{(n)}(\ubk_N)}}\\
\nonumber
&\bigg(\prod_{u_1\in J_1}a^\dagger(\bk_{u_1})\bigg)
\bigg(\prod_{v_1\in I_n\backslash J_1}a(\bk_{v_1})\bigg)\dotsm\\
\label{SK}
&\dotsm\bigg(\prod_{u_N\in J_N}a^\dagger(\bk_{u_N})\bigg)
\bigg(\prod_{v_N\in I_n\backslash J_N}a(\bk_{v_N})\bigg)
\end{align}
where
\begin{eqnarray}
\nonumber
\lefteqn{K_\lambda^{J_\bullet}(\ubk_1,\dotsc,\ubk_N)=}\\
\nonumber
&=&c_{n,N}\int_{{\mathbb R}^{4nN}}d\ua_1\dotsm d\ua_N\;
\bigg(\prod_{1\leq M< N}\theta\big(\kappa^0(\ua_{M+1}^0-\ua_M^0)\big)\bigg)
G_\lambda(\ua_1)\dotsm G_\lambda(\ua_N)\\
\label{K}
&&
\exp\left\{i\sum_{M=1}^N{\uk_M}_\mu\cdot\big({U_M}{\ua_M}\big)^\mu\right\}.
\end{eqnarray}
Here ${U_M}$ is a diagonal $n\times n$ matrix,
with diagonal entries
\begin{equation}
\label{U}
{U_M}_{uu}=
\begin{cases}
        \phantom{-}1,\quad u\in J_M,\\
        -1,\quad u\in I_n\backslash J_n,
\end{cases}
\end{equation}
and
\[
\tilde\uk=(\tilde k_1,\dotsc,\tilde k_M).
\]

By the second Wick theorem, each term in the above sum may be written
as a sum of terms with contractions, where the surviving creation
 and destruction operators appear in normal order (creation on the left), and
the contracted pairs are replaced by $\delta^{(3)}$ distributions. Since
each integration variable ${\bk_M}_j$ appears as the argument of
precisely one operator
$a^\sharp$, ($a^\sharp=a,a^\dagger$), then the arising product 
of $\delta$s is well defined by the following elementary
\begin{remark} The map
\label{good}
\begin{equation}
\label{goodintegral}
f\mapsto
\int da_1\dotsm da_d f(a_1,\dotsm,a_d)\prod_{j=1}^s\delta(l_j(a_1,\dotsc,a_d)),
\quad s\leq d,
\end{equation}
on the Schwartz functions, with $l^j$ a real linear functional 
on $\mathbb R^d$, $j=1,\dotsc,s$,
is a well defined distribution
if and only if  the functionals $l^1,\dotsc,l^s$ are linearly independent.
By performing $s$ integrations, the above distribution always takes the form
\[
\int db_1\dotsm db_{d-s}\;f\big(a_1(b_1,\cdots,b_{d-s}),
\dotsc,a_d(b_1,\cdots,b_{d-s})\big),
\]
where the linear maps $a_j=a_j(b_1,\cdots,b_{d-s})$, $j=1,d$, 
provide a linear injection of $\mathbb R^{d-s}$ into $\mathbb R^d$.
\end{remark}
After complete Wick reduction, the generic term 
will be of the form
\[
\int_{\mathbb R^{6k}}d\bp_1\dotsm d\bp_{r+s}W(\bp_1,\dots,\bp_{r+s})
a^\dagger(\bp_1)\dotsm a^\dagger(\bp_r)
a(\bp_{r+1})\dotsm a(\bp_{r+s}),
\]
where
\begin{eqnarray*}
\lefteqn{W(\bp_1,\dotsc,\bp_{r},\bp_{r+1},\dotsc,\bp_{r+s})=}\\
&=&\frac{K_\lambda^{J_\bullet}(\ubk_1(\bp_1,\dotsc,\bp_{r+s}),\dotsc,
\ubk_N(\bp_1,\dotsc,\bp_{r+s})}%
{\sqrt{\omega^{(n)}(\ubk_1(\bp_1,\dotsc,\bp_{r+s}))
\dotsm\omega^{(n)}(\ubk_N(\bp_1,\dotsc,\bp_{r+s}))}}
\end{eqnarray*}
for a suitable set of linear maps (see remark \ref{good}) 
\[
\ubk_1=\ubk_1(\bp_1\dotsm\bp_{r+s}),\dotsc,\ubk_N=\ubk_N(\bp_1\dotsm\bp_{r+s})
\]
with $r+s\leq nN$.\footnote{Of course, the case $r+s=nN$ corresponds to the 
term with no contractions; in that case, the above mentioned 
linear maps reduce to renaming the integration variables.}
The proof is complete  by \cite[Theorem X.44]{reed_simon2},
if we show that the function $W$ is in $L^2(\mathbb R^{3(r+s)})$.
To this end, it is enough to prove that
$|K_\lambda^{J_\bullet}|$ is bounded by a Schwartz function. Due to
the form of $\lambda$,
the integrations over the space and time variables in (\ref{K}) factorize,
and $K_\lambda^{J_\bullet}$ is the product of a function with Gaussian decay 
with the Fourier transform~---~evaluated at some point continuously
depending on the $\ubk_M$'s~---~of an $L^1$ function; the latter is then a 
bounded, continuous function. Indeed, we will
show in appendix \ref{lambdasecondoproof} that $K^{J_\bullet}_\lambda$
is a Schwartz function, by carrying out explicitly the above mentioned 
computations.\qed

Note that the   kernels
$W(\bp_1,\dotsc,\bp_{r},\bp_{r+1},\dotsc,\bp_{r+s})$ appearing in the above
proof are 
transition amplitudes of scattering processes with $s$ incoming and
$r$ outgoing particles. They decay as
Gaussian functions of the energies and momentum transfers
expressed in Planck units.

\section{Time-ordering and Diagrammatics}
\label{diagramm}

We complement our discussion with some formal remarks aiming to clarify the
relation between the approach followed here and the usual formulation of the
perturbation theory in terms of Feynman  diagrams.  We
have already shown in~\cite{bdfp1} 
that the perturbative setup dealt with
in~\cite{dfr} (as well as the approach based on the Yang-Feldman equation) is
inequivalent to the  by now standard setup  in terms of the modified Feynman
rules~\cite{filk}. In this section we will make explicit that in the
framework of the regularized interaction proposed in this paper, Feynman
propagators are no more available at all.   

In this discussion we ignore all problems which may arise in the adiabatic
limit, and use the formal  (i.e. defined with constant coupling constant
$\lambda$)  time dependent
interaction Hamiltonian $H_I(t)$ throughout,
\[
H_I(t)=\lambda\int_{\mathbb R^{4n}} dx_1\dotsm dx_n\;w_t(x_1,\dotsc,x_n)
\wick{\phi(x_1)\dotsm\phi(x_n)},
\]
where the integral kernel $w_t$ is of the form
\[
w_t(x_1,\dotsc,x_n)=w'(x_1,\dotsc,x_n)\,
\delta^{(1)}\big(t-\frac{1}{n}\sum_j {x_j}^0\big),
\]
and the Gaussian kernel $w'$ does not depend on~$t$.
Introducing the time ordered kernels
\[
\mathcal T_{t_1,\dotsc,t_N}=
\sum_{\pi\in\mathbb P_N}
\theta \big(t_{\pi(1)}-t_{\pi(2)}\big)\dots
\theta\big(t_{\pi(N-1)}-t_{\pi(N)}\big)
w_{t_{\pi(1)}}\otimes\dots\otimes w_{t_{\pi(n)}},
\]
formal computations yield the formal Dyson expansion (see \cite{dyson})
\begin{eqnarray*}
\nonumber
\lefteqn{S=I+\sum_{N=1}^{\infty}\frac{(-i\lambda)^N}{N!}
\int dt_1\dots dt_N\;
\int d^{4n}\underline x_1\dots d^{4n}\underline x_N}\\
&&
\mathcal T_{t_1,\dots,t_N}(\underline x_1,\dots \underline x_N)
\wick{\phi^n(\underline x_1)}\dots
\wick{\phi^n(\underline x_N)},
\end{eqnarray*}
where we use the notations (\ref{Smart1}--\ref{Smart8}).

By integrating over the time variables $t_j$, we obtain
\begin{eqnarray*}
\lefteqn{S=I+\sum_{N=1}^{\infty}\frac{(-i\lambda)^N}{N!}
\int d^{4n}\underline x_1\dots d^{4n}\underline x_N}\\
&&
w(\underline x_1)\dotsm w(\underline x_N)
AT\wick{\phi^n(\underline x_1)}\dots
\wick{\phi^n(\underline x_N)}.
\end{eqnarray*}
Here $AT$
is the ordering of the Wick monomials with respect to the average
time, i.e.~the time of the mean position of each $\underline{x}_j$: more 
precisely, with
\begin{equation}
\label{atau}
\tau(\underline x)=\kappa^0(\ux^0)=\frac{x_1^0+\dotsb+x_n^0}{n}
\end{equation}
the time component of the four vector 
$(x_1+\dotsb+x_n)/n$, 
we have
\begin{eqnarray*}
\lefteqn{
AT\wick{\phi^n(\underline x_1)}\dotsm\wick{\phi^n(\underline x_N)}\equiv}\\
&\equiv&\sum_{\pi\in\mathbb P_n}\theta\big(\tau(\underline x_{\pi(1)})-
\tau(\underline x_{\pi(2)})\big)\dotsm
\theta\big(\tau(\underline x_{\pi(N-1)})-\tau(\underline x_{\pi(N)})\big)\\
&&\wick{\phi^n(\underline x_{\pi(1)})}\dotsm
\wick{\phi^n(\underline x_{\pi(N)})},
\end{eqnarray*}
where $\underline x_j=(\underline x_{j,1},\dotsc,\underline x_{j,n})$ can be
thought of as a ``fat vertex'',   i.e. a multi-vertex  
actually consisting of $n$ Minkowski vectors.

The time ordering thus defined, as well as the one discussed in~\cite{dfr},
arises from ordering the variables $t_1,\dotsc,t_N$ in products
$H_I(t_1)\dotsm H_I(t_N)$.

However, in both approaches causality is violated at
the Planck length scale   (cf. also~\cite{grosse}).
A typical manifestation of the violation of causality is the following:
suppose that  
\[
AT\wick{\phi^n(\underline x)}\wick{\phi^n(\underline y)}=
\wick{\phi^n(\underline x)}\wick{\phi^n(\underline y)},
\] 
i.e.
$\tau(\underline x)<\tau(\underline y)$, then   it is not 
forbidden that ${x_i}^0>{y_{i^\prime}}^0$ for some $i,i^\prime\in\{1,\dots,n\}$. 
In other words, a single field
$\phi(x_i)$ belonging to the first Wick monomial can be subsequent in time
to some field $\phi(y_{i^\prime})$.

Note however, that the above picture is consistent with the request that the
large scale limit should reproduce the ordinary (non regularized) theory with
$\wick{\phi^n}$ interaction; in particular, as $\lambda_P\rightarrow 0$, the
multi-vertices shrink to their mean positions, and the above average
time-ordering reduces to the usual one.

We conclude our discussion by showing that it is not possible to absorb the
time ordering in Feynman propagators, $i\Delta_F(x-y) = 
(\Omega,T[\phi(x),\phi(y)]\Omega)$, as one usually does in the framework of
ordinary local theories. In fact, in the approach followed here, we cannot
construct Feynman propagators {\em at all}, while in the original Hamiltonian
approach based on~(\ref{old_wick}) as well as in the approach based on the
Yang-Feldman equation it is still possible to rewrite the time ordering in
terms of Feynman propagators {\em together with} advanced and retarded
propagators~\cite{bdfp1}. 

\begin{figure}
\begin{centering}
\epsfig{file=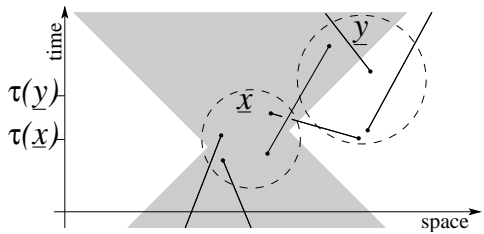, height=3cm}
\caption{\footnotesize{
Multi-vertices are represented by the
points within a circle; $\tau(\underline x)$ is the
average time of the multi-vertex~$\underline x$, and the shadowed area
represents its causal completion.}}
\label{qfish}
\end{centering}
\end{figure}

To see what happens in the approach adopted here, 
consider as an example the second order contribution to the Dyson
series  with two internal contractions in quantum $\phi^4$-interaction, as
depicted in figure~\ref{qfish}. 

Here, the two multi-vertices are sets of four distinguishable points,
\linebreak
$\{x_1,x_2,x_3,x_4\}$ and $\{y_1,y_2,y_3,y_4\}$, and at most one
line originates from each of these points. A
kernel $w'$ belongs to each multi-vertex. The 
internal contractions lead to positive and negative frequency parts of the
commutator function,
$\Delta_\pm(x_i-y_{i^\prime})$, where $i,i^\prime \in \{1,\dots,4\}$. , while the
time ordering is done with respect to the time components of the mean
coordinates of each vertex, leading to the Heaviside functions
$\theta\left(\pm\left((x_1^0+\dots+x_4^0)- (y_1^0+\dots+y_4^0)\right)\right)$.
We may thus conclude that the distributions arising in the fish graph will not 
yield Feynman propagators at all. 

This is consistent with the fact that the usual interpretation of Feynman
graphs as pictorial representations of scattering processes (``first A is
annihilated, then B is created'') should break down on quantum spacetime, as it
needs sharply localized events, which are no longer available.

\appendix

\section*{Appendix}
\label{lambdasecondoproof}
In this appendix, we explicitly compute the kernel $K^{J_\bullet}_\lambda$, 
for future reference.

Recall that , for any $\bk\in\mathbb R^3$, we write 
$\tilde k=(\omega(\bk),\bk)$, and $\tilde\uk=(\tilde k_1,\dotsc,\tilde k_n)$.
Let $f\in\mathcal S(\mathbb R^{3nN})$ be any Schwartz function, and 
$\lambda(t,\bx)=\lambda'(t)\lambda''(\bx)$. 
By equation (\ref{K}), 
\[
\langle K_\lambda^{J_\bullet},f\rangle=\int_{\mathbb R^{3nN}}
d\ubk_1\dotsm d\ubk_N\;
F_\lambda^{J_\bullet}(\tilde\uk_1,\dotsc,\tilde\uk_N)
f(\ubk_1,\dotsc,\ubk_N).
\]
where $F_\lambda^{J_\bullet}\in
\mathcal{C}(\mathbb R^{4nN})$ is given by
\begin{eqnarray*}
\lefteqn{F_\lambda^{J_\bullet}(\uk_1,\dotsc,\uk_N)=}\\
&=&c_{n,N}\int_{{\mathbb R}^{4nN}}d\ua_1\dotsm d\ua_N\;
\bigg(\prod_{1\leq M< N}\theta(\kappa^0(\ua_{M+1}^0-\ua_M))\bigg)
G_\lambda(\ua_1)\dotsm G_\lambda(\ua_N)\\
&&\exp\left\{i\sum_{M=1}^N{\uk_M}_\mu\cdot\big({U_M}{\ua_M}\big)^\mu\right\};
\end{eqnarray*}
the orthogonal matrix ${U_M}$ is given by (\ref{U}).

We now set 
\[
\uxi=n^{-1/2}(1,1,\dotsc,1)\in\mathbf R^{n};
\]
Then, if $P$ is the orthogonal projection of $\mathbb R^n$ onto 
$\mathbb R\uxi$, we set $P'=(P,\boldsymbol P)$ as a projection onto 
$\mathbb R^{4n}$, $P'=P\oplus P\oplus P\oplus P$;
In the same spirit, $\uxi'=
(\uxi,\underline{\boldsymbol{\xi}})\in\mathbb R^{4n}$. Finally,
$U'_M=({U_M},\boldsymbol {U_M})$, and 
$I'=(I,\boldsymbol I)$ is the identity. We finally need the maps
$V_\uxi(\ua)=\uxi\cdot\ua$, $V_{\ue_n}(\ua)=\ue_n\cdot\ua$ from $\mathbb R^n$
to $\mathbb R$, with the corresponding maps 
$V'_\uxi=(V_\uxi,\boldsymbol V_\uxi)$,
$V'_{\ue_n}=(V_{\ue_n},\boldsymbol V_{\ue_n})$ from $\mathbb R^{4n}$ to 
$\mathbb R^4$.

With these notations, we have
\begin{gather*}
\kappa^0(\ua^0)=\frac{\uxi\cdot\ua^0}{\sqrt n}=
\frac{V_{\uxi}(\ua^0)}{\sqrt n},\\
\kappa(\ua)=\frac{V'_{\uxi}(\ua^0)}{\sqrt n},\quad
\boldsymbol\kappa(\uba)=\frac{\boldsymbol V_{\uxi}(\uba)}{\sqrt n}.
\end{gather*}
Moreover, the function $G_\lambda$ defined by (\ref{GSchwartz})
may be written as
\[
G_\lambda(\ua)=
\lambda\bigg(\frac{V'_{\uxi}\cdot\ua}{\sqrt n}\bigg)
\exp\bigg\{-\frac{1}{2}|(I'-P')\ua|^2\bigg\},
\]
and
\begin{eqnarray*}
\lefteqn{F_\lambda^{J_\bullet}(\uk_1,\dotsc,\uk_N)=}\\
&=&c_{n,N}\int_{{\mathbb R}^{4nN}}d\ua_1\dotsm d\ua_N\;
\bigg(\prod_{1\leq M< N}\theta(\uxi\cdot(\ua_{M+1}^0-\ua_M))\bigg)
\left(\prod_{M=1}^N\lambda\bigg(\frac{V'\uxi\cdot\ua_M}{\sqrt n}\right)\\
&&\exp\bigg\{-\frac{1}{2}\sum_{M=1}^N|(I'-P')\ua_M|^2\}
\exp\left\{i\sum_{M=1}^N{\uk_M}_\mu\cdot\big({U_M}{\ua_M}\big)^\mu\right\};
\end{eqnarray*}

We may now take a $n\times n$ real orthogonal matrix $R$, such that
\[
R\uxi=\ue_n=(0,0,\dotsc,0,1)\in\mathbf R^{n}.
\]
Then, $E=R^tPR$ is the orthogonal projection of $\mathbb R^{n}$
onto $\mathbb R\ue_n$, and $E'=(E,\boldsymbol E)$, as   usually.
Furthermore, 
$V'_{e_n}(R\ua)=V'_{\uxi}(\ua)$.
With the change of integration variables $\ubb=R\uba$, we obtain
\begin{eqnarray*}
\lefteqn{F_\lambda^{J_\bullet}(\uk_1,\dotsc,\uk_N)=}\\
&=&c_{n,N}\int_{{\mathbb R}^{4nN}}d\ub_1\dotsm d\ub_N\;
\bigg(\prod_{1\leq M< N}\theta(\ue'_n\cdot(\ub_{M+1}^0-\ub_M))\bigg)
\left(\prod_{M=1}^N\lambda\bigg(\frac{V'_{\ue_n}(\ub_M)}{\sqrt n}\bigg)\right)\\
&&\exp\bigg\{-\frac{1}{2}\sum_{M=1}^N|(I'-E')\ub_M|^2\}
\exp\left\{i\sum_{M=1}^N{\big(R{U_M}\uk_M}_0\big)\cdot\ub_M^0\right\}\\
&&\exp\left\{-i\sum_{M=1}^N{\big(\boldsymbol{RU}_M\ubk_M}\big)\cdot
\ubb_M\right\};
\end{eqnarray*}

The integration over the variables ${\ub_M}_n\in\mathbb R^4$, 
$M=1,\dotsc,N$, 
is completely separated from the integration over the variables
$\ub_M^*=({b_M}_1,\dotsc,{b_M}_{n-1})\in\mathbb R^{4(n-1)}$. Hence 
\begin{eqnarray*}
\lefteqn{F_\lambda^{J_\bullet}(\uk_1,\dotsc,\uk_N)=}\\
&=&c_{n,N}\left(\int_{{\mathbb R}^{4(n-1)N}}d\ub_1^*\dotsm d\ub_N^*
\exp\bigg\{-\frac{1}{2}\sum_{M=1}^N|\ub_M^*|^2\bigg\}\right.\\
&&\left.\exp\left\{-i\sum_{M=1}^N{\big(\boldsymbol{RU}_M\ubk_M}\big)^*\cdot
\ubb_M^*\right\}
\exp\left\{i\sum_{M=1}^N{\big(R{U_M}\uk_M}_0\big)^*\cdot{\ub_M^0}^*\right\}
\right)\times\\
&&\times\bigg(\int_{\mathbb R^{4N}} d{b_1}_n\dotsm d{b_N}_n\;
\bigg(\prod_{1\leq M< N}\theta(\ue_n\cdot({{b_{M+1}}_n}^0-{{b_M}_n}^0))\bigg)
\left(\prod_{M=1}^N\lambda\bigg(\frac{{{b_M}_n}^0}{\sqrt n}\bigg)\right)\\
&&
\exp\left\{i\sum_{M=1}^N\big(V'_{e_n}(R{U_M}\uk_M\big)_\mu{{\ub_M}_n}^\mu\right\}\bigg).
\end{eqnarray*}
The first factor is a Fourier transform of a Gaussian function.  
Renaming the integration variables 
${b_M}_n=\beta_M=(\beta_M^0,\boldsymbol\beta_M)\in\mathbb R^4$, 
\begin{eqnarray*}
\lefteqn{F_\lambda^{J_\bullet}(\uk_1,\dotsc,\uk_N)=}\\
&=&\frac{c_{n,N}}{(2\pi)^{4(n-1)N}}
\exp\bigg\{-\frac{1}{2}\sum_{M=1}^N
\big|(I'-P')U'_M\uk_M\big|^2\bigg\}\times\\
&&\times\bigg(\int_{\mathbb R^{4N}} d{\beta_1}\dotsm d{\beta_N}\;
\bigg(\prod_{1\leq M< N}\theta(\beta_{M+1}^0-{\beta_M^0}))\bigg)
\left(\prod_{M=1}^N\lambda\bigg(\frac{\beta_M}{\sqrt n}\bigg)\right)\\
&&
\exp\left\{i\sum_{M=1}^N\big(V'_{\uxi}({U_M}\uk_M)\big)_\mu\beta_M^\mu\right\}\bigg),
\end{eqnarray*}
where we used
\[
\big|\big(R'{U'_M}\uk_M\big)^*\big|=
\big|(I'-E')R'{U'_M}\uk_M\big|=
\big|(I'-P')U'_M\uk_M\big|.
\]

Due to the form of $\lambda=\lambda'\otimes\lambda''$, the integral in the
above expression may be further factorized  as an integral over the variables 
$\bbeta_1,\dotsc,\bbeta_N$, times an integral over the variables 
$\beta_1^0,\dotsc,\beta_N^0$:
\begin{eqnarray*}
\lefteqn{F_\lambda^{J_\bullet}(\uk_1,\dotsc,\uk_N)=}\\
&=&\frac{c_{n,N}}{(2\pi)^{4(n-1)N}}
\exp\bigg\{-\frac{1}{2}\sum_{M=1}^N
\big|(I'-P')U'_M\uk_M\big|^2\bigg\}\times\\
&&\times\left\{\int_{\mathbb R^{3N}} d{\bbeta_1}\dotsm d{\bbeta_N}\;
\left(\prod_{M=1}^N\lambda''\bigg(\frac{\bbeta_M}{\sqrt n}\bigg)\right)
\exp\left\{-i\big(\boldsymbol V_{\uxi}({\boldsymbol U_M}
\ubk_M)\big)\cdot\bbeta_M\right\}.
\right\}\times\\
&&\times\left\{\int_{\mathbb R^{N}} d{\beta_1^0}\dotsm d{\beta_N^0}\;
\exp\left\{i\big(V_{\uxi}({U_M}\uk_M^0)\big)\beta_M^0\right\}
\left(\prod_{M=1}^N\lambda'\bigg(\frac{\beta^0_M}{\sqrt n}\bigg)\right)\times\right.\\
&&\times\left.
\bigg(\prod_{1\leq M< N}\theta(\beta_{M+1}^0-{\beta_M^0}))\bigg)\right\}
\end{eqnarray*}

The integral over $\beta_1^0,\dotsc,\beta_M^0$ in the above expression
is the Fourier transform of an $L^1$ function, hence it 
vanishes at infinity. Moreover, it is the product of a Schwartz function
in $\ubk_1,\dotsc,\ubk_N$, times a $\mathcal C^\infty$ function
of ${\uk_1}_0,\dotsc,{\uk_N}_0$, vanishing at infinity.
As a consequence, 
\[
(\ubk_1,\dotsc,\ubk_N)\mapsto 
K^{J_\bullet}_\lambda(\ubk_1,\cdots,\ubk_N)=
F^{J_\bullet}_\lambda(\tilde\uk_1,\cdots,\tilde\uk_N)
\]
is a Schwartz function.


\end{document}